# A Coherent Beam Splitter for Electronic Spin States

J. R. Petta[1], H. Lu[2], A. C. Gossard[2]

[1] Department of Physics, Princeton University, Princeton, NJ 08544, USA.

[2] Materials Department, University of California at Santa Barbara, Santa Barbara, CA 93106, USA.

Rapid coherent control of electron spin states is required for implementation of a spin-based quantum processor. We demonstrate coherent control of electronic spin states in a double quantum dot by sweeping an initially prepared spin singlet state through a singlet-triplet anti-crossing in the energy level spectrum. The anti-crossing serves as a beam splitter for the incoming spin singlet state. Consecutive crossings through the beam splitter, when performed within the spin dephasing time, result in coherent quantum oscillations between the singlet state and a triplet state. The all-electrical method for quantum control relies on electron-nuclear spin coupling and drives single electron spin rotations on nanosecond timescales.

Energy level crossings, where two quantum states cross in energy as a function of an external parameter, are ubiquitous in quantum mechanics (*1*). Coupling of the quantum states provided by tunnel coupling with strength Δ, for example, leads to hybridization of the states and results in an anti-crossing with a minimum energy splitting 2Δ (*2,3*). Passing a quantum state through an anti-crossing in the level diagram will result in a sweep-rate-dependent non-adiabatic transition probability, $P_{LZ}$, commonly known as the Landau-Zener probability (*4*). The theory of Landau-Zener transitions can be applied to a diverse set of problems, ranging from electronic transitions in molecular collisions, to chemical reactions, to neutrino conversion in the sun (5). We apply Landau-Zener transition physics to coherently control electronic spin states in a semiconductor double quantum dot (DQD).

Semiconductor quantum dots have emerged as promising platforms for quantum control of charge and spin degrees of freedom (*6*). Considering future applications of electron spin qubits in quantum information processing, the required elementary building blocks are the exchange gate, which couples two spins, and single spin rotations (*7*). Extremely fast 200 picosecond exchange gates have been demonstrated (*6,8*). However, coupling to the small magnetic moment of the electron (as required for single spin rotations) is much more difficult, leading to relatively long ~100 ns gate operation times in GaAs quantum dots (*9*). In addition, the ac magnetic fields required for single spin electron spin resonance (ESR) are difficult to localize on a single quantum dot (~40 nm), hindering extension of the method to a large number of quantum dots operating in close proximity. Several groups have demonstrated fast optical control of single spins, but these methods are also difficult to apply locally (*10,11*). In principle, local rotations can be achieved using electrically driven spin resonance (EDSR), which requires spin-orbit coupling and an ac electric field, but the Rabi frequencies obtained in GaAs quantum dots are approximately a factor of two slower than those obtained using conventional ESR (*12,13*). We demonstrate an all-electrical method for driving local single spin rotations on nanosecond timescales.

Our method for coherent quantum control of electron spins is based on two consecutive sweeps through a singlet-triplet anti-crossing in a DQD energy level diagram. Coherent oscillations between the singlet and $m_s$=+1 triplet state, $T_+$, occur on a nanosecond timescale and are made possible by the hyperfine interaction between the trapped electron spins and the nuclear spin bath (*14-16*). The oscillations are controlled by tuning the external magnetic field, $B_E$, and the voltage pulse profile that sweeps the quantum dot system through the anti-crossing in the energy level diagram. Similar sweeps through energy level anti-crossings in superconducting qubits have been used to study Landau-Zener interference (*17-21*). In addition, deeply bound molecular states have been generated by transferring weakly bound Feshbach molecules through a series of anti-crossings in a molecular energy manifold (*22*).

In our device (Fig. 1A) depletion gates are arranged in a triple quantum dot geometry (*23*). A DQD is formed using the middle and right dots of the device. Gate voltages $V_L$ and $V_R$ are used to tune the device to the (1,1)-(2,0) charge transition, where ($N_L$, $N_R$) indicate the number of electrons in the (left, right) dot. High sensitivity charge sensing is achieved by depleting gates $Q_1$ and $Q_2$ to form a quantum point contact (QPC) charge sensor with conductance $g_Q$ (*8*). Energy level anti-crossings (Fig. 1B) in the DQD can be used for quantum control in a manner that is directly analogous to an optical beam splitter (*18-20*).

The detuning, ε, of the DQD (Fig. 2A) is adjusted using gate voltages $V_L$ and $V_R$ (*24*). For positive detuning the ground state is the spin singlet (2,0)S. By decreasing the detuning, a single electron can be transferred from the left dot to the right dot, forming a (1,1) charge state. Here the possible spin state configurations are the spin singlet, S, and the spin triplets $T_0$, $T_-$, and $T_+$ with $m_S$=0,-1, and +1 respectively. (2,0)S and S hybridize near ε=0 due to the interdot tunnel coupling, $T_c$. The $T_+$ and $T_-$ states are separated from the $T_0$ state by the Zeeman energy, $E_Z=g\mu_B(B_E+B_N)$, where $B_N$ is the Overhauser field ($B_N^{rms}$~2 mT in the absence of nuclear polarization) (*15*). Throughout this work we take $|g|$=0.44, based on previous experiments (*8,25*). We focus on the boxed region in Fig. 2A, where hyperfine interactions mix the S and $T_+$ states resulting in an anti-crossing in the energy level diagram. Under appropriate experimental conditions we show that this anti-crossing functions as a beam splitter for incoming quantum states (*18-20*).

We first measure the quantum state transition dynamics at the S-$T_+$ avoided crossing in order to verify the mechanism of Landau-Zener tunneling. The analytical expression for the non-adiabatic transition probability is $P_{LZ} = e^{-\frac{2\pi\Delta^2}{\hbar\nu}}$ (*4*). Here ħ is Planck's constant divided by 2π and ν is the energy level velocity, defined as $\nu=|d(E_1-E_2)/dt|$, where $E_1$ and $E_2$ are the energies of the states involved in the anti-crossing. We determine Δ by measuring $P_{LZ}$ as a function of the sweep rate through the S-$T_+$ anti-crossing. A (2,0)S state is first prepared at positive detuning, then a rapid gate voltage pulse (~1.1 ns, non-adiabatic with respect to the S-$T_+$ mixing rate) shifts the system to negative detuning, $\varepsilon_S$, which preserves the spin singlet, S. The detuning is then increased during a ramp time $T_R$, sweeping the system back through the S-$T_+$ avoided crossing. A QPC charge sensor determines the final singlet state probability, $P_S$, via spin-to-charge conversion (*6*).

$P_S$ is plotted in Fig. 2B as a function of the ramp time $T_R$. For long ramp times the initial state should follow the adiabatic branch during the return sweep through the S-$T_+$ anti-crossing, resulting in a final state $T_+$, as illustrated in the inset of Fig. 2B. We measure $P_S$~0.3 at long $T_R$, due to the limited measurement contrast set by the spin relaxation time. At short times $P_S$ decays exponentially as expected from the Landau-Zener model, with a characteristic timescale of ~180 ns. Given the detuning pulse amplitude, 1.7 mV, and the conversion between gate voltage and energy, $|d(E_S-E_{T+})/d\varepsilon|$~3.9 μeV/mV, we extract a best fit Δ=60 neV (*24*). In comparison, time resolved measurements of the S-$T_+$ spin dephasing time yield $T_2$*=10 ns, corresponding to an energy scale of 66 neV, which is in good agreement with the value of Δ obtained above (*8*). In superconducting flux qubits this tunnel splitting is set by tunnel junction parameters, whereas in the S-$T_+$ qubit Δ is set by fluctuating transverse hyperfine fields (*15,18*).

Quantum control of the S and $T_+$ states is achieved by consecutively passing through the S-$T_+$ avoided crossing in the coherent limit, where the consecutive crossings take place within the spin dephasing time (*18-21*). The opposite limit, where $T_R$>>$T_2$*, has been shown to lead to dynamic nuclear polarization (*26*). Our pulse sequence for quantum control is illustrated in Fig. 3A and is analogous in operation to an optical interferometer (inset, Fig. 3C). An initially prepared spin singlet state is swept through the S-$T_+$ avoided crossing. During this detuning sweep, the S-$T_+$ avoided crossing "splits" the incoming singlet state into a superposition of states S and $T_+$, with amplitudes $A_S$ and $A_{T+}$, analogous to an optical beam splitter. In correspondence with the Landau-Zener equation, $|A_S|^2=P_{LZ}$. Spin angular momentum is conserved during this process by coupling to the nuclear spin bath via the hyperfine interaction, resulting in a small amount of nuclear polarization (*16,26*). Detuning is then maintained at a value $\varepsilon_S$ for the nominal pulse length $\tau_S$, which results in a phase accumulation $\phi = \frac{1}{\hbar}\int\left[E_S\left(\varepsilon(t)\right) - E_{T+}(\varepsilon(t))\right]dt$ that is equivalent to changing the path length of one leg of an optical interferometer. A second detuning sweep takes the system back through the S-$T_+$ anti-crossing resulting in quantum interference of the two paths. The singlet state return probability, $P_S$, is measured using the QPC charge sensor.

The consecutive sweeps through the S-$T_+$ anti-crossing and the intermediate phase accumulation, $\phi$, can be treated as unitary operations (Fig. 3B) that act on the initially prepared spin singlet state (*4,27,28*). For the ideal case of $P_{LZ}$=1/2, the S-$T_+$ anti-crossing functions as a 50:50 beam splitter resulting in the unitary operator $U_1 = {}^{1}\!/_{\sqrt{2}}(\sigma_x - \sigma_z)$, which is equivalent to a Hadamard gate. Phase accumulation, $\phi$, during the detuning pulse results in a $\sigma_Z$ rotation, $U_2 = exp(\frac{-i}{2}\phi\sigma_z)$, while the return sweep back through the S-$T_+$ anti-crossing in the limit $P_{LZ}$=1/2 results in a third unitary operation $U_3 =$ $1/\sqrt{2}\,(\sigma_x - \sigma_z)$. Functional forms for the unitary operators under general driving conditions are given in the supporting online material (*24*).

The measured $P_S$ shows clear Stückelberg oscillations between S and $T_+$ as a function of $\tau_S$ and $\varepsilon_S$ (*4,5*). At negative detunings, far from the avoided crossing, the oscillation period is set by $E_S$-$E_{T+}$=$E_Z$. For $B_E$=100 mT, the Zeeman energy corresponds to a period of 1.6 ns assuming $|g|$=0.44, in good agreement with the ~1.5 ns period observed in the data for $\varepsilon_S$=-1.7 mV. The curvature of the interference pattern is partially due to the voltage pulse profile, which is smoothed to maintain some degree of adiabaticity during the sweep through the S-$T_+$ anti-crossing. In these data (Fig. 3C) the first peak corresponds to the condition where the detuning pulse exactly reaches the S-$T_+$ anti-crossing. The second interference fringe corresponds to a configuration in which $U_2$ gives a π-pulse about the z-axis of the Bloch sphere.

Singlet state probability as a function of pulse length, $P_S(\tau_S)$, is plotted in Fig. 3D for two different values of detuning. The oscillation visibility ranges from 15% to 30% for these data and is a function of detuning since the spin relaxation time and $P_{LZ}$ are detuning dependent (*6*). Higher visibility oscillations are obtained when the level velocity, ν, is small at the S-$T_+$ anti-crossing (insets to Fig. 3D). Maximum visibility would be obtained for $P_{LZ}$=1/2 (the limit of a perfect 50:50 beam splitter). To achieve this, detuning ramp times on the order of 160 ns>>$T_2$* are required, which is no longer in the coherent limit. These data suggest that active pulse shaping with sub-nanosecond resolution could be used to increase the fidelity of the gate operations by lowering the level velocity only in the vicinity of the S-$T_+$ anti-crossing.

We confirm that the interference fringes are caused by consecutive sweeps through the S-$T_+$ beam splitter by varying the external field, $B_E$. A reduction in $B_E$ lowers $E_S$-$E_{T+}$ and shifts the position of the S-$T_+$ anti-crossing to more negative values of ε, both of which reduce the phase accumulation during $U_2$ for a fixed set of voltage pulse parameters (Fig. 4A). Landau-Zener interference patterns are plotted in Figs. 4B-D for $B_E$=90, 70, and 50 mT. A reduction in field results in two major differences: 1) the first oscillation shifts to more negative $\varepsilon_S$ and 2) the oscillation frequency decreases. Both observations are consistent with the level diagram shown in Fig. 4A.

To quantitatively model the data we calculate the probability to return to the spin singlet state, $P_S$, by considering the action of the unitary operations (Fig. 3B) on the initially prepared spin singlet state. Neglecting relaxation and dephasing, we find $P_S = 1 - 2P_{LZ}(1 - P_{LZ})[1 + \cos(\phi - 2\tilde{\phi}_S)]$, where $\tilde{\phi}_S$ is related to the Stoke's phase (*19,24*). We calculate the accumulated phase $\phi$ by combining our knowledge of the voltage pulse profile with the measured $E_S(\varepsilon)$-$E_{T+}(\varepsilon)$, as determined by energy level spectroscopy (*24*). The visibility of the calculated oscillations (see insets in Fig. 4B-D) is 15% and is set by $P_{LZ}$=0.96, as determined for these sweep conditions using the data in Fig. 2B. Overall, the observed and calculated Landau-Zener interference patterns are in very good agreement. The decay of the oscillations as a function of $\tau_S$ is most likely due to fluctuations in the Overhauser field (*8*).

While commonly used single spin rotation mechanisms rely on gigahertz frequency magnetic fields, the coherent rotations between S and $T_+$ demonstrated here occur on a nanosecond timescale set by the Zeeman energy and are solely driven using local gate voltage pulses. As a result, it will be feasible to scale this quantum control method to a large number of spin qubits operating in close proximity. In addition, it is possible that the spin-flip mechanism employed here, which relies on coupling to the nuclear spin bath, could be harnessed under the appropriate conditions to create a nuclear spin memory (*29*).

30. We acknowledge useful discussions with Guido Burkard, Bill Coish, Duncan Haldane, David Huse, Daniel Loss, and Hugo Ribeiro and thank Chris Laumann for technical contributions. Research at Princeton was supported by the Sloan Foundation, the Packard Foundation, and the National Science Foundation through the Princeton Center for Complex Materials, DMR-0819860, and CAREER award, DMR-0846341. Work at UCSB was supported by the DARPA grant N66001-09-1-2020 and the UCSB National Science Foundation DMR MRSEC.

**Figures**

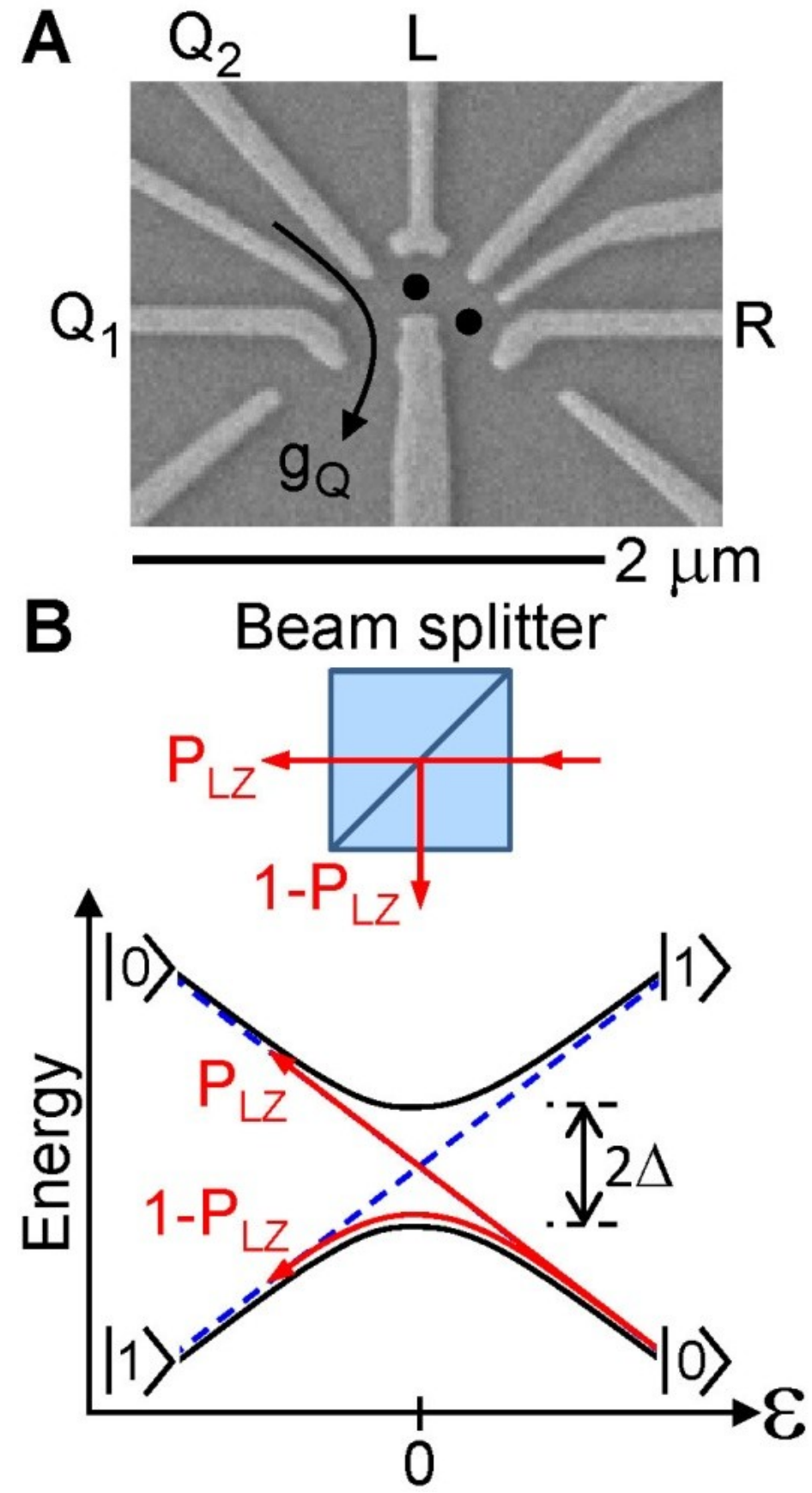

**Fig. 1. (A)** Scanning electron microscope image of a device similar to the one used in this experiment. Voltages on gates L and R tune the occupation of the DQD, while gates $Q_1$ and $Q_2$ form a QPC with conductance, $g_Q$, for single charge sensing. **(B)** Energy level anti-crossings can be used to "split" an incoming quantum state, in direct analogy with an optical beam splitter. The non-adiabatic transition probability, $P_{LZ}$, depends on the level velocity, ν, and the energy splitting at the anti-crossing, 2Δ.

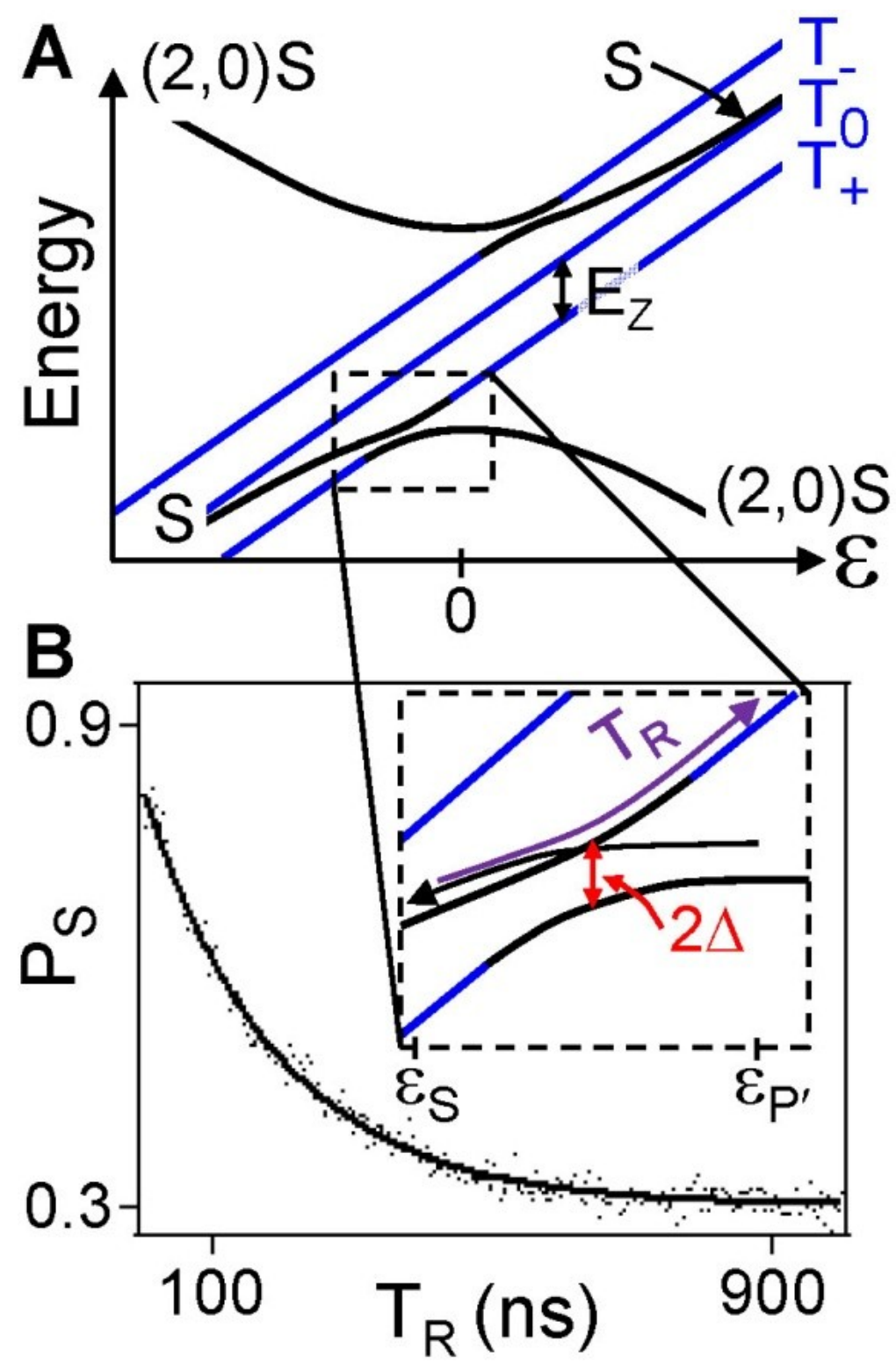

**Fig. 2. (A)** DQD energy level diagram near the (1,1)-(2,0) charge transition. Hyperfine fields result in an anti-crossing between the S and $T_+$ states (dashed box), which serves as a beam splitter for quantum control. **(B)** The Landau-Zener transition probability, $P_{LZ}$, is measured by preparing (2,0)S at positive detuning and then converting it to S via a rapid, ~1.1 ns, gate voltage detuning pulse from $\varepsilon_{P'}$ to $\varepsilon_S$. The detuning is then increased at a constant rate from $\varepsilon_S$ to $\varepsilon_{P'}$ during a time interval $T_R$ and a QPC measures the final singlet state probability, $P_S$. The data are fit to an exponential decay (solid line) as expected from the Landau-Zener transition formula, resulting in a best fit coupling strength Δ=60 neV.

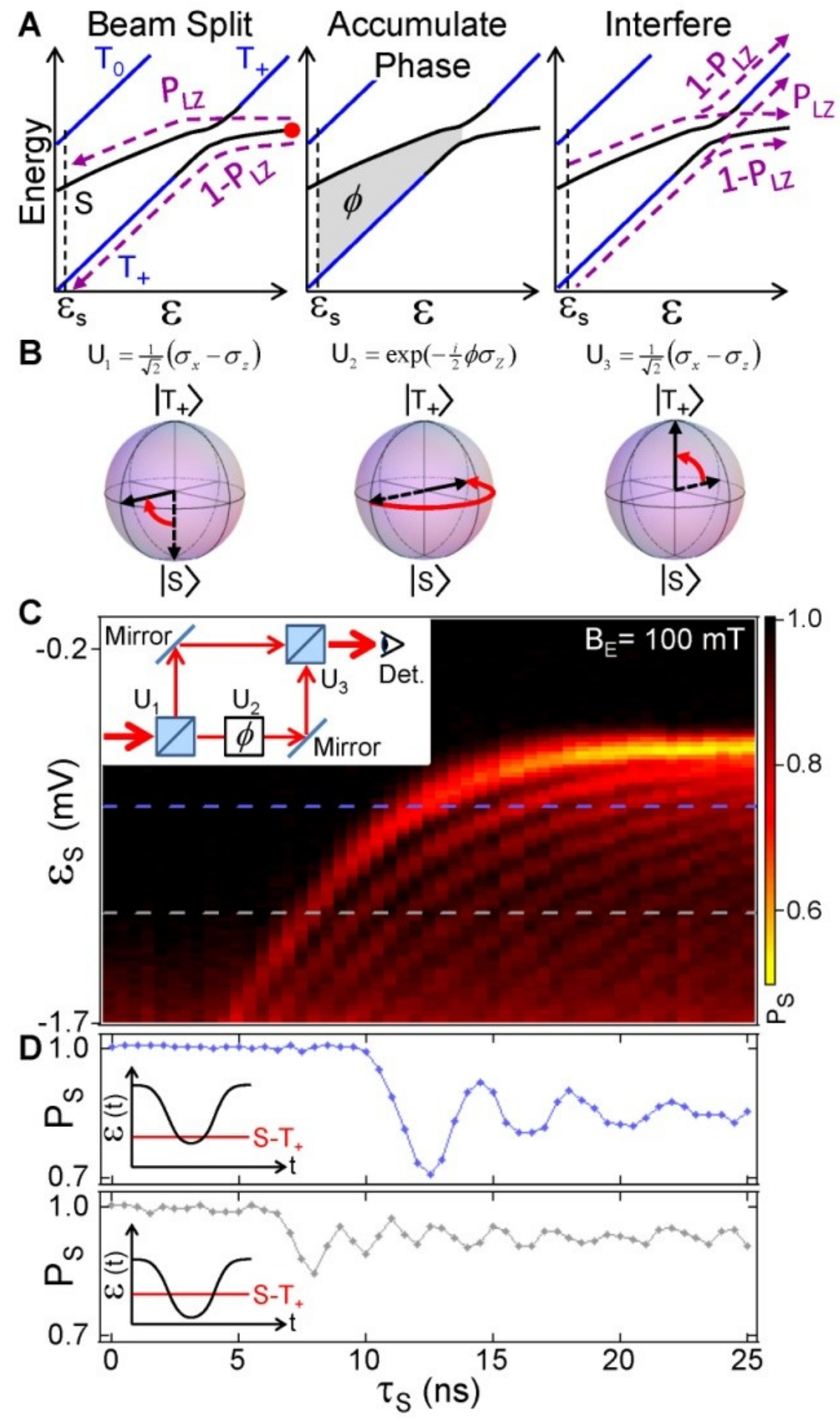

**Fig. 3.** **(A)** From left to right: An initially prepared (2,0)S state is swept through the S-$T_+$ anti-crossing, resulting in a superposition of states S and $T_+$, with amplitudes $A_S$ and $A_{T+}$, analogous to an optical beam splitter. The energy difference between these two states results in relative phase accumulation, $\phi$, which can be controlled by tuning $B_E$ and the gate voltage pulse profile. A return sweep through the S-$T_+$ anti-crossing results in quantum interference and the final state is determined using spin to charge conversion. **(B)** Bloch sphere representation of the unitary rotations for specific sweep conditions resulting in $P_{LZ}$=1/2 and $\phi$=π. **(C)** The singlet state return probability, $P_S$, displays

coherent oscillations as a function of separation detuning, $\varepsilon_S$, and pulse length, $\tau_S$, due to Landau-Zener interference. Inset: The experiment is equivalent to an optical interferometer, where a change in path length of one of the interferometer arms results in interference fringes as observed by a detector (Det.). **(D)** Singlet state probability as a function of pulse length, $P_S(\tau_S)$, extracted from the data in **(C)** for two different values of $\varepsilon_S$.

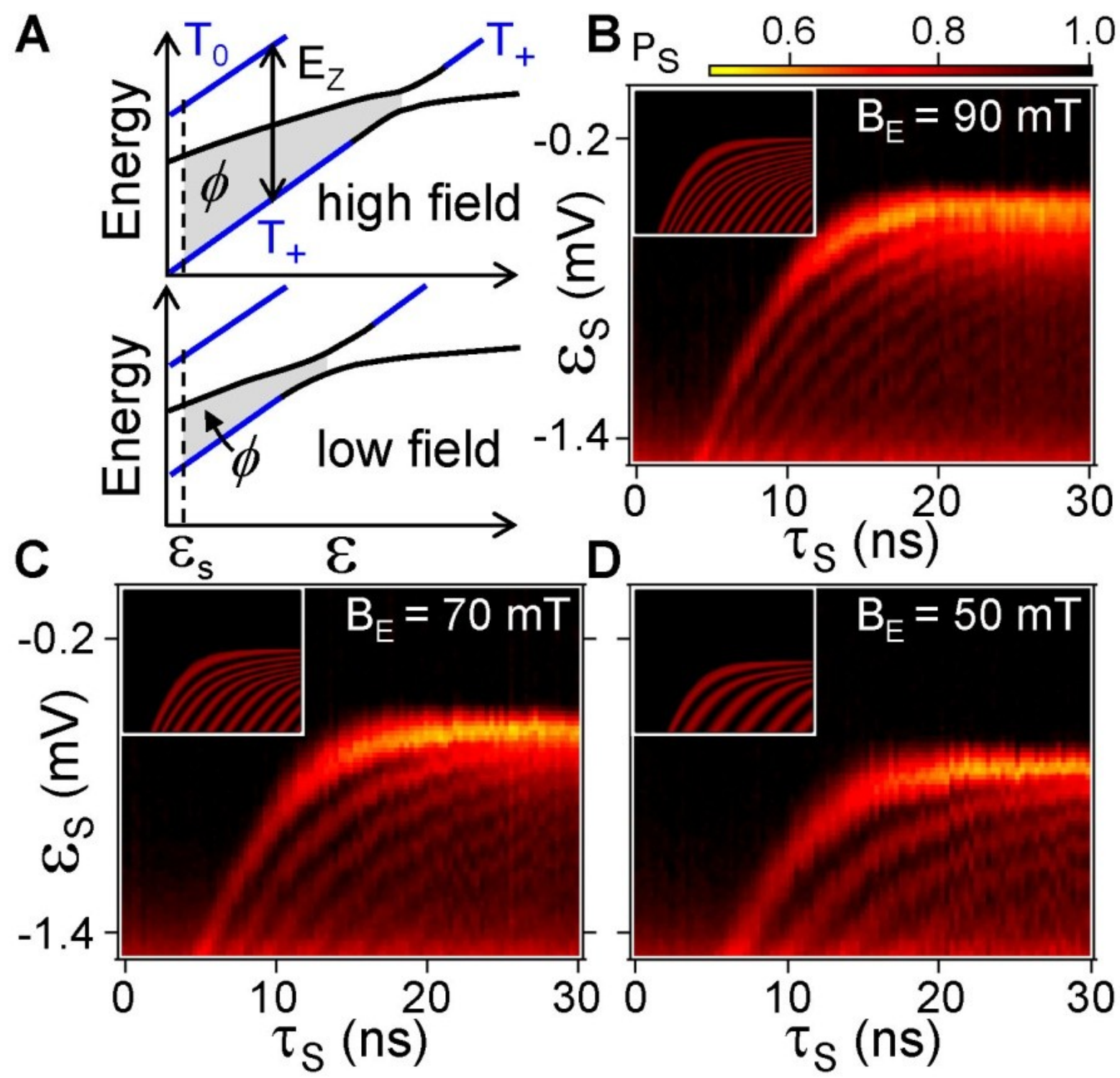

**Fig. 4. (A)** The accumulated phase, $\phi$, is controlled by tuning $B_E$ for a fixed set of voltage pulse parameters. A reduction in $B_E$ shifts the position of the S-$T_+$ anti-crossing to more negative values of $\varepsilon_S$ and reduces $E_S$-$E_{T+}$. **(B)** , **(C)** , and **(D)** Measured Landau-Zener interference patterns for $B_E$=90, 70, and 50 mT, respectively. Insets: calculated interference patterns (see text).